\documentclass[aps,prb,twocolumn,superscriptaddress]{revtex4-1}

\usepackage[pdftex]{graphicx}
\def \etal {{\textit {et al.}\ }}

\begin{document}


\title{Observation of Multiple-Gap Structure in Hidden Order State of URu$_2$Si$_2$ from Optical Conductivity}


\author{Jesse S. Hall}

\affiliation{Department of Physics and Astronomy, McMaster University, Hamilton, ON L8S 4M1, Canada}

\author{Travis Williams}
\affiliation{Department of Physics and Astronomy, McMaster University, Hamilton, ON L8S 4M1, Canada}

\author{Graeme Luke}
\affiliation{Department of Physics and Astronomy, McMaster University, Hamilton, ON L8S 4M1, Canada}
\affiliation{The Canadian Institute for Advanced Research, Toronto, Ontario M5G 1Z8, Canada}

\author{Urmas Nagel}
\affiliation{National Institute of Chemical Physics and Biophysics, Akadeemia tee 23, 12618 Tallinn, Estonia}

\author{Taaniel Uleksin}
\affiliation{National Institute of Chemical Physics and Biophysics, Akadeemia tee 23, 12618 Tallinn, Estonia}

\author{Toomas R\~o\~om}
\affiliation{National Institute of Chemical Physics and Biophysics, Akadeemia tee 23, 12618 Tallinn, Estonia}

\author{Thomas Timusk}
\affiliation{Department of Physics and Astronomy, McMaster University, Hamilton, ON L8S 4M1, Canada}
\affiliation{The Canadian Institute for Advanced Research, Toronto, Ontario M5G 1Z8, Canada}


\date{\today}

\begin{abstract}
We have measured the far infrared reflectance of the heavy fermion compound URu$_2$Si$_2$ through the phase transition at T$_{HO}$=17.5 K dubbed 'hidden order' with light polarized along both the a- and c-axes of the tetragonal structure. The optical conductivity allows the formation of the hidden order gap to be investigated in detail. We find that both the conductivity and the gap structure are anisotropic, and that the c-axis conductivity shows evidence for a double gap structure, with $\Delta_{1,c}=2.7$ meV and $\Delta_{2,c}=1.8$ meV respectively at 4 K, while the gap seen in the a-axis conductivity has a value of $\Delta_a=3.2$ meV at 4 K. The opening of the gaps does not follow the behaviour expected from mean field theory in the vicinity of the transition.
\end{abstract}


\maketitle


\section{ Introduction} 

The heavy fermion compound URu$_2$Si$_2$ has been extensively studied due to the great diversity of electronic behaviours it displays in different temperature regimes. At room temperature, URu$_2$Si$_2$ behaves as a poor metal due to the Kondo effect, with a Kondo temperature of approximately \cite{schoenes} 370 K and a slowly increasing resistivity as the temperature is lowered, reaching a maximum of $\sim 320$ $\mu\Omega$cm at about 70 K. As the temperature is lowered further the resistivity begins to decrease.\cite{maple, palstra2, mcelfresh, zhu} This decrease has been attributed to the formation of a Kondo lattice where the localized uranium f-electrons hybridize with the conduction electrons forming heavy charge carriers with $m^*\sim25 m_e$ in this material, \textit{i.e.} heavy Fermions. \cite{palstra2,maple} This view has been challenged recently by Schmidt \etal \cite{schmidt} using STM and confirmed by Nagel \etal \cite{nagel} using infrared spectroscopy, who find that the mass of the charge carriers remains on the order of the free electron mass down to 17.5 K.  At this temperature URu$_2$Si$_2$ undergoes a second-order phase transition, identifiable as a discontinuity in the resistivity \cite{maple, palstra2, mcelfresh, zhu} and specific heat, \cite{palstra} for which no order parameter has yet been conclusively identified. 

The specific heat data fit well to an exponential decay below the ordering temperature as would be expected in a BCS-like transition, with a corresponding loss of entropy of approximately 0.2 R ln 2. The DC electrical resistivity data fit well to a Fermi liquid model above the transition, revealing that electron-electron interactions are the dominant scattering mechanism in the coherent regime. Below the transition DC resistivity fits well to an exponential decay with an additional $T^2$ term to account for the continued contribution to the scattering from Fermi liquid physics. There is a pronounced anisotropy between the a-direction and the c-direction, with the c-direction having a lower resistivity. 

Initially, the transition was assumed to be antiferromagnetic ordering \cite{schlabitz, palstra} or the onset of a  density wave state \cite{maple}. Magnetic ordering was subsequently ruled out by neutron scattering \cite{broholm} which detected an ordered moment of 0.03 $\mu_B$ per U atom, too small to account for the entropy loss inferred from the specific heat. The alternative of a charge density wave state has received little support since there is no evidence of lattice distortion. Because the order parameter remains unknown the transition has been named hidden order. A number of theoretical models have been proposed to account for the order parameter, such as multipolar ordering, density and hybridization waves \cite{kiss,chandra,haule,cricchio,harima}, but so far none has been conclusively identified as responsible for the transition.

Previous reflectance measurements \cite{bonn, degiorgi} have studied the structure and evolution of the frequency-dependent conductivity, measured in the a-b cleavage plane as a function of temperature. Above 75 K the optical conductivity is completely frequency independent, characteristic of incoherent hopping conductivity. Below 75 K the spectrum is well described as a metal, with a Drude peak that becomes sharper and narrower down through the hidden order transition. More recently, measurements on an a-c face \cite{crazy} above the hidden order transition have demonstrated the anisotropy in the optical conductivity expected from transport measurements. Further measurements on the a-b plane \cite{nagel} have recently shown that the spectral weight associated with the Drude peak is constant between 75 K and $T_{HO}$, suggesting that the effective mass of ~$5$ $m_e$ remains constant between these temperatures, and that major changes to the electronic structure do not occur in this temperature range.

We have conducted optical spectroscopy measurements on oriented samples of URu$_2$Si$_2$ along both the a-axis and c-axis, using a new technique for obtaining high quality, low-noise data. We present the first comparison of optical spectra from both crystal directions on URu$_2$Si$_2$ in the hidden order state. 

\section{Experimental Method} 

Single crystals of  URu$_2$Si$_2$ were grown at McMaster University by the Czochralski method. The crystals were grown in a tri-arc furnace in argon atmosphere and were then either cleaved perpendicular to the c-axis or oriented with a Laue x-ray diffractometer and cut with a tungsten wire saw parallel to the c-axis. The ac-face was then polished smooth and etched using a wash of HF acid to remove any damaged surface after it was found that the polishing process removed the phonon features from the reflectance. 

The DC resistivity has been measured along the a-axis on crystals from the same growth as those used in the reflectance measurements, and along both axes elsewhere \cite{palstra, mcelfresh, zhu}. The form of the resistivity is robust between different samples, and the values of the resistivity are close enough to one another that it seems more likely that the differences are due to uncertainty in the positioning of the electrodes and the sample geometry during the measurements than variations due to sample quality. When the resistivity at the transition is normalized to agree between the DC measurements taken at McMaster, on a crystal grown at the Institute N\'eel, and in reference \cite{zhu}, the values along the a-axis agree with one another to better than 10 \%  up to 30 K. We therefore use the DC conductivity from Zhu \etal \cite{zhu} for the c-axis.

The samples were measured with standard reflectance techniques, using both an immersion cryostat with a $^3$He bolometer and a Sciencetech SPS spectrometer in Tallinn, Estonia, and an open-flow helium cryostat and commercial Bruker IFS 66 v/s spectrometer at McMaster. A standard gold overcoating technique \cite{homes} was used to get absolute spectra accurate to 0.3\% and reproducible between the two measurement systems. 

In order to interpret the electronic behavior as the gap forms across the Fermi surface at the transition, it is necessary to extract the optical conductivity from the absolute reflectance using the Kramers-Kronig relations. As a first approximation, since  URu$_2$Si$_2$ behaves electronically as a metal at low temperatures both above and below the hidden order transition, the Hagen-Rubens formula for the reflectance in the limit of low frequencies, $R(\omega)=1-\sqrt{(\frac{2\omega}{\pi\sigma_{DC}})}$, has been used to extrapolate the reflectance to zero frequency.  Measured DC conductivity was used to compare expected reflectance to that measured. In this case, the measured absolute reflectance was adjusted slightly to agree with the transport data, as slight drifts in the measurement system, particularly the detectors used, can cause errors on the order of 0.5\%.

To eliminate geometrical artifacts at low frequency the sample was held stationary during the measurement process and the temperature was varied. Ratios were then constructed between a reference temperature above the HO transition and different temperatures below it containing very little noise. Geometrical artifacts are introduced to the absolute reflectance during the gold overcoating process due to the motion of the sample, detector drift, and the imperfect reproduction of the original sample position. The noise in the absolute spectra is greater than the noise in the temperature ratios by an order of magnitude. In order to address this, a simple polynomial can be fit to the absolute reflectance at the reference temperature. This smoothed estimate of the absolute reflectance at the reference temperature and the ratios are then combined to give absolute reflectance at the temperatures of interest. This removes weak temperature independent features if they are within the noise level, but allows the transition to be studied with only the much smaller noise from the temperature ratios. 

\section{Results} 

\begin{figure}
\includegraphics[width=3.5 in]{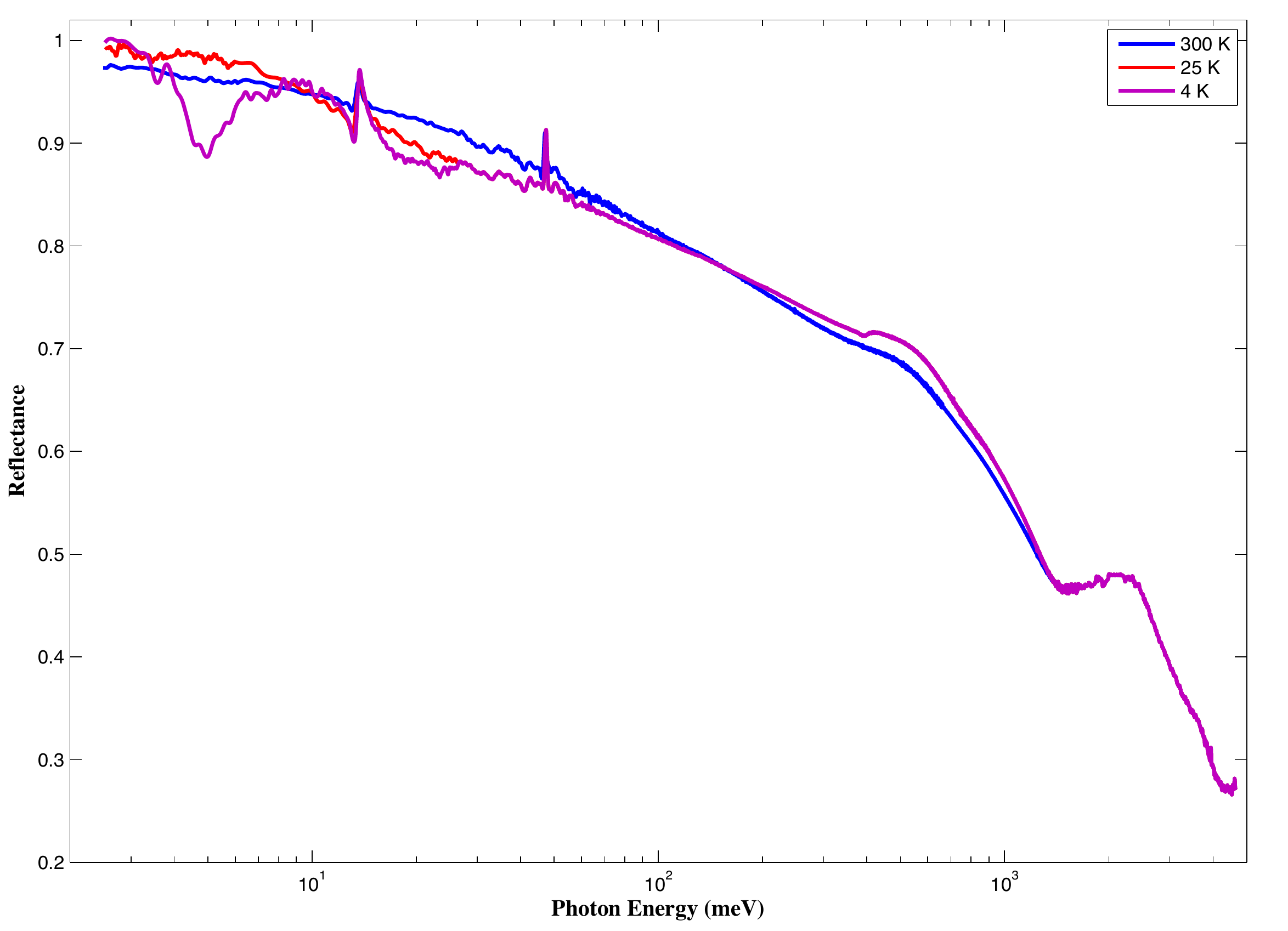}%
\caption{\label{spectrum}Absolute reflectance measured in the ab-plane from the far infrared to the ultraviolet. The shoulder-like features at 380 meV and 1200 meV correspond to interband transitions. The partial hybridization gap appears in the 25 K spectrum as a drop in the reflectance between 15 and 30 meV. The Hidden Order gap opening causes a strong absorption centred around 5 meV, visible in the 5 K spectrum.}
\end{figure}

The full reflectance in the ab plane is show in Figure \ref{spectrum}. There are optical phonons at 13.6 and 47.1 meV in the a-axis and 42.8 meV in the c-direction. Interband transitions can be seen as a shoulder-like feature at 380 meV and 1200 meV. The feature at 380 meV has a sharper onset at lower temperatures that broadens out at room temperature. The signature of the hidden order can be seen as a strong absorption at 5 meV at the lowest temperature, and it is the only feature within our spectral range that can be associated with the hidden order transition. The drop in reflectance between room temperature and 25 K in the region between 12 meV and 40 meV can be attributed to the formation of a hybridization pseudogap \cite{nagel, crazy} below 70 K.

\begin{figure}
\includegraphics[width=3.5in]{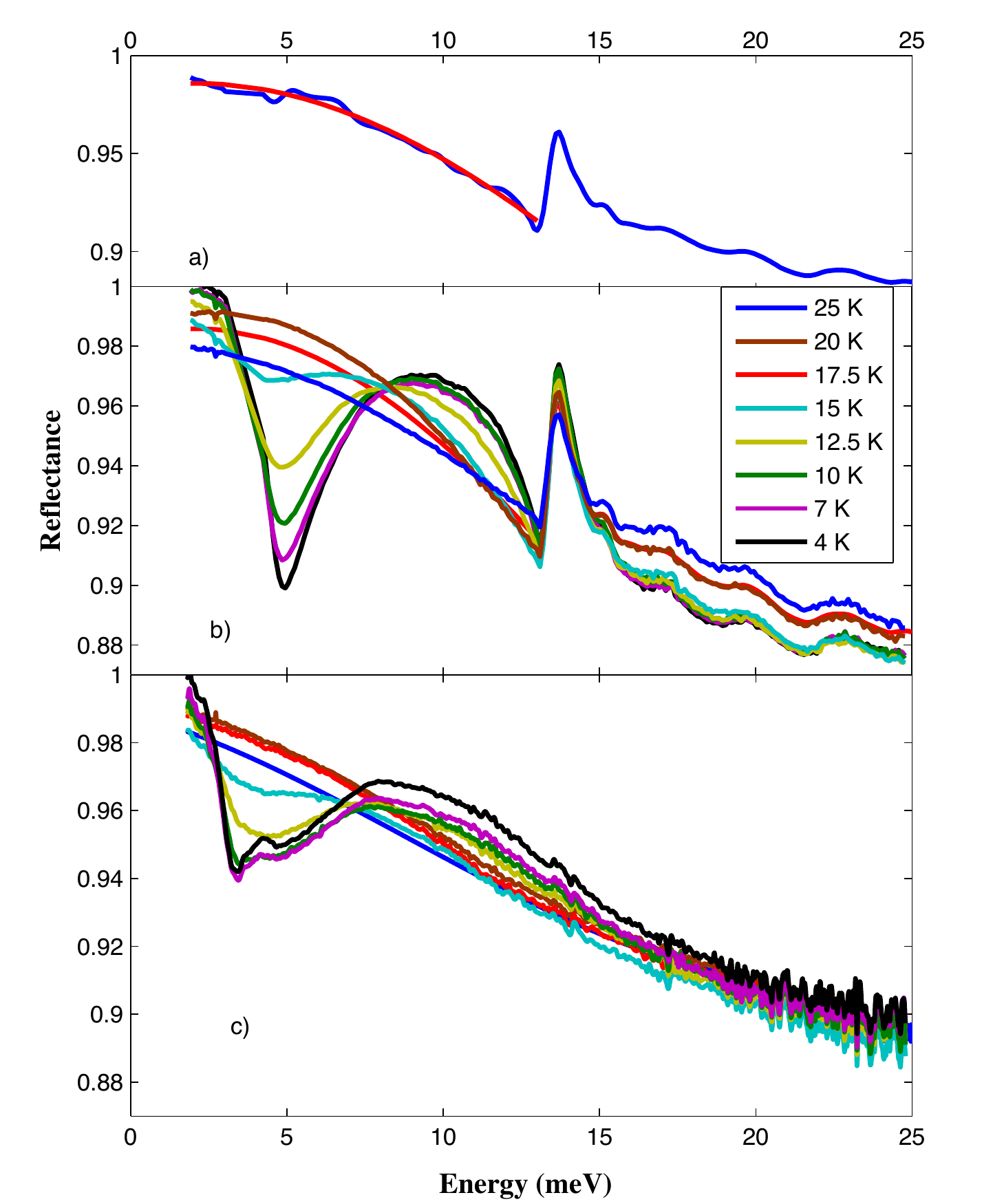}%
\caption{\label{reflectivity} a) Absolute reflectance in the a-axis at 25 K, with the polynomial fit to smooth out geometrical artifacts introduced during motion of the sample. The same procedure was used for the c-axis (not shown). b) Refined reflectance along the a-axis, obtained by multiplying the fit to the 25 K absolute reflectance with the measured reflectance ratios at different temperatures. c) Refined reflectance along the c-axis. At the lowest temperatures, additional structure appears within the absorption around the minimum.}
\end{figure}

Figure \ref{reflectivity} shows the polynomial fit to the absolute reflectance described above, along with the smoothed absolute reflectance along both distinct crystal directions. The onset of the hidden order state is clearly visible in the low temperature spectrum as a drop in the reflectance of both crystal axes centered around 5 meV (a-axis) and 4 meV (c-axis). The minimum is distinct and shifts to slightly higher energies with decreasing temperature. The appearance of this minimum evolves gradually with a sudden onset at the hidden order temperature and its depth increases monotonically with decreasing temperature in the a-direction; in the c-direction there is additional structure that makes this impossible to determine.  In the c-axis below 12 K an additional feature appears in the absorption near the minimum, unlike when the light is polarized along the a-axis whose absorption minimum remains sharp to our lowest temperature of 4 K. No other sharp features appear below T$_{HO}$ above 2 meV.

\begin{figure}
\includegraphics[width=3.5in]{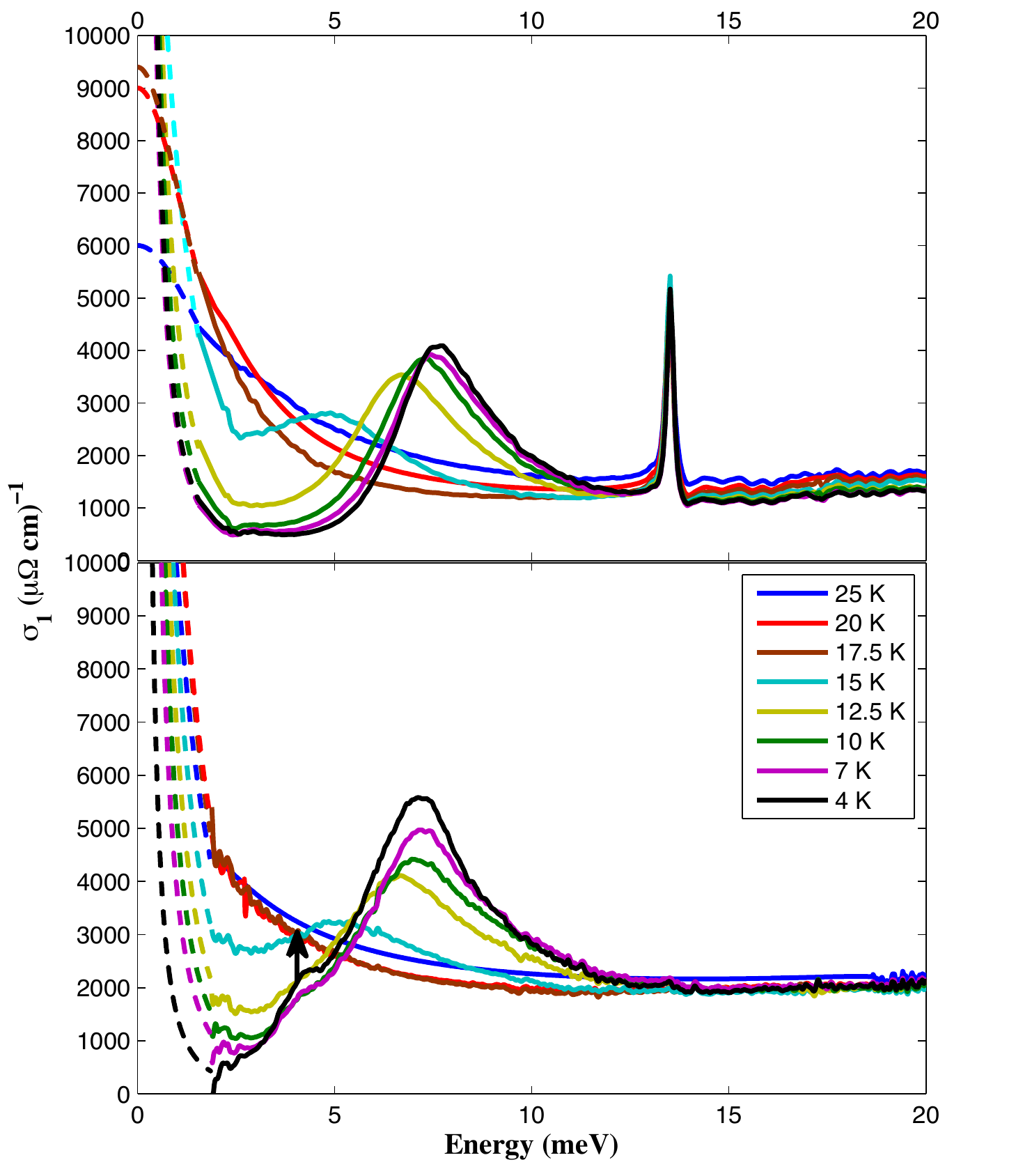}%
\caption{\label{conductivity}Real part of the optical conductivity of the a-axis (upper panel) and c-axis (lower panel) for selected temperatures above and below the transition. The arrow in the lower panel indicates the position of the second gap at lower temperatures. The dashed lines at low frequency indicate the Drude peak that has been extrapolated to agree with DC resistivity measurements.}
\end{figure}

 Figure \ref{conductivity} shows the optical conductivity at selected temperatures. The solid curves above 2 meV show the real part of the optical conductivity from the Kramers-Kronig analysis as described above. URu$_2$Si$_2$ behaves electronically as a metal in the low temperature regime both above and below $T_{HO}$. To obtain an estimate of the conductivity below 2 meV shown as dashed curves, where we have no optical data, we used the following procedure to fit a Drude peak to the conductivity. The amplitude of the peak $\sigma(0)$ was taken from the measured dc conductivity and the width $1/\tau$ from the tail of the Drude peak that extended to the optically measured region above 2 meV.

As expected from transport data, there is a strong anisotropy between the conductivity in the a-direction and the c-direction. Both have qualitatively the same features above the transition: a sharp Drude peak and strong incoherent background, with the Drude peak becoming sharper as coherence becomes stronger with decreasing temperature. The conductivity is higher along the c-axis, consistent with DC resistivity measurements. 

The gap can be identified as a suppression of the conductivity in the frequency range immediately above the narrow Drude peak, with a shift of spectral weight to the frequencies in a narrow region above the gap energy, visible as a sharp peak. The Drude peak narrows sharply and increases in height as the gap develops, but the spectral weight lost in the gap region is not fully recovered either in the Drude peak or above the gap. The very strong suppression of the conductivity in the gap region suggests that the gap forms across a large section of the Fermi surface. Qualitatively, the gap structure appears similar in the two crystal directions, although the peak above the gap is broader in the c-direction and the gap energy is larger in the a-direction. The structure within the absorption in the reflectance along the c-axis is visible within the gap region in the optical conductivity as an additional bump appearing below 12 K, becoming stronger and sharper as the temperature decreases. 

\begin{figure}
\includegraphics[width=3.5in]{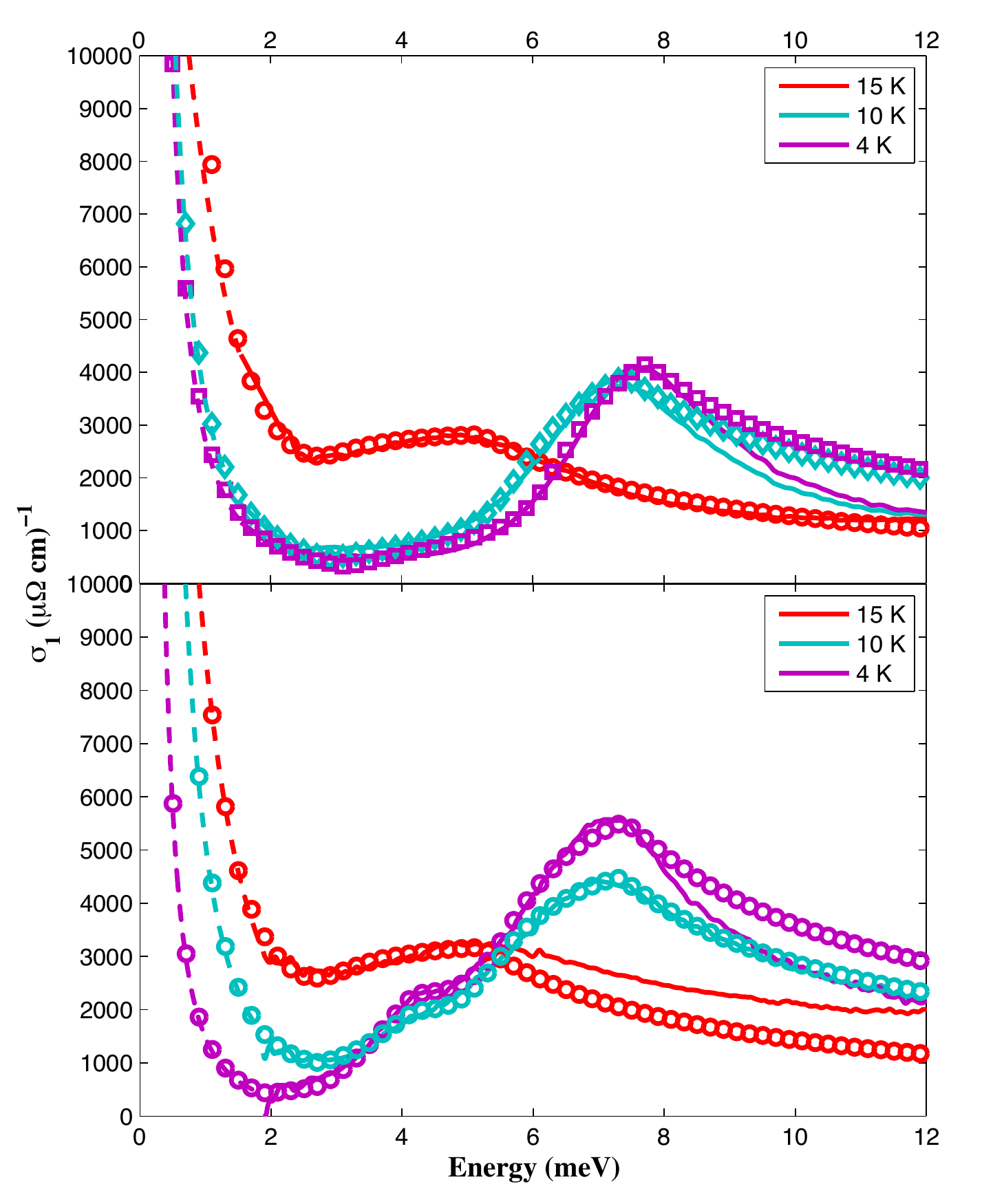}%
\caption{\label{gap_fits}Fits of a Dynes model for the density of states to the optical conductivity for the a-axis (top panel) and c-axis (bottom panel) for selected temperatures in the hidden order state. The dashed lines at low frequency indicate the fit to a Drude model. Above the cut-off frequency, exact agreement is not expected since the details of the bandstructure are not taken into account in the fitting process.}
\end{figure}

We add an isotropic s-wave gap model from Dynes \etal \cite{dynes} with a square-root-like singularity in the density of states to attempt to parameterize the gap seen in the conductivity: 
\begin{eqnarray} 
n_D(E)=|Re\frac{E/\Delta + i\gamma}{\sqrt{(E/\Delta + i\gamma)^2-1}}| 
\end{eqnarray} 

\noindent where $\Delta$ is the gap energy, $\gamma$ is a damping parameter, and a cutoff frequency $\omega_c$ is introduced to account for the region where the density of states goes to zero. The comparison of the isotropic gap model fit to the measured conductivity is shown in Figure \ref{gap_fits} for selected temperatures for both crystal directions. The behavior of the relevant bandstructure above the cutoff is unknown and the form of the cutoff is somewhat arbitrary and has not been included in the model, so exact agreement above this frequency is not necessarily to be expected. 

The value of $\gamma$ is determined by two factors: the width of the gap and the quasiparticle lifetime. In both polarizations and for both gaps in the c-direction $\gamma$ is constant with temperature. The width of the gap was determined to be the same for all three gaps to within 10\%. The quasiparticle liftime is the same for the a-axis and larger c-axis gaps, but differs for the smaller c-axis gap, leading to an overall variation in the value of $\gamma$. The values for $\gamma$ due to the quasiparticle lifetime are $\gamma_{a,qp} = \gamma_{c1,qp} = 0.1 meV$, $\gamma_{c2,qp} = 0.5 meV$.

The appearance of multiple gaps, as well as the differences between the gap sizes, observed with different axes of polarization indicates that the gaps are not isotropic. The isotropic gap model still works extremely well, however, suggesting that the density of states does indeed possess a square-root singularity despite the anisotropy of the underlying gap. The structure of the Fermi surface is complicated with as many as five separate sheets \cite{mydoshrev} which makes a reasonable determination of gap structure difficult. Optical measurements are by their nature averages over all of k-space in the direction given by the light polarization, so variations in the detailed gap structure will not necessarily produce greatly differing optical conductivity.

The a-axis gap evolves steadily down into the hidden order state, with a value at 4 K of $\Delta_a=3.1$ meV in good agreement with ARPES \cite{santander} (which sees a heavy band dispersing approximately 4 meV below the Fermi level) and tunneling \cite{aynajian} ($\Delta=4.1$ meV), as well as previous optical data \cite{bonn}. The gap in the c-axis is not easy to compare to other measurements because the material does not cleave along the ac-face, so surface-sensitive techniques like STM and ARPES cannot be used directly. Our measurements suggest that the additional structure in the c-axis conductivity is due to a second gap opening within the hidden order state. The larger gap has a value of $\Delta_{1,c}=2.7$ meV, while the smaller gap value is $\Delta_{2,c}=1.8$ meV, both at 4 K. The isotropic gap model can be fit at all temperatures without changing the quasiparticle widths or lifetimes or the gap width; only the value of the gap needs to be changed to fit the data, except at 15 K where the cutoff frequency has to be decreased slightly. 

The gap in the a-axis is well described by the single-gap s-wave model fit, while the c-axis requires a two-gap model to be consistent with the data. The second gap seen in the c-axis is too weak to be seen in the 15 K conductivity, but is clearly present at the 12.5 K and is nearly fully developed before it is visible, with a value of $\Delta_2=1.5 $meV at T=12.5 K and $\Delta_2=1.8 $meV at T=4 K. 

\section{Discussion} 

\begin{figure}
\includegraphics[width=3.5in]{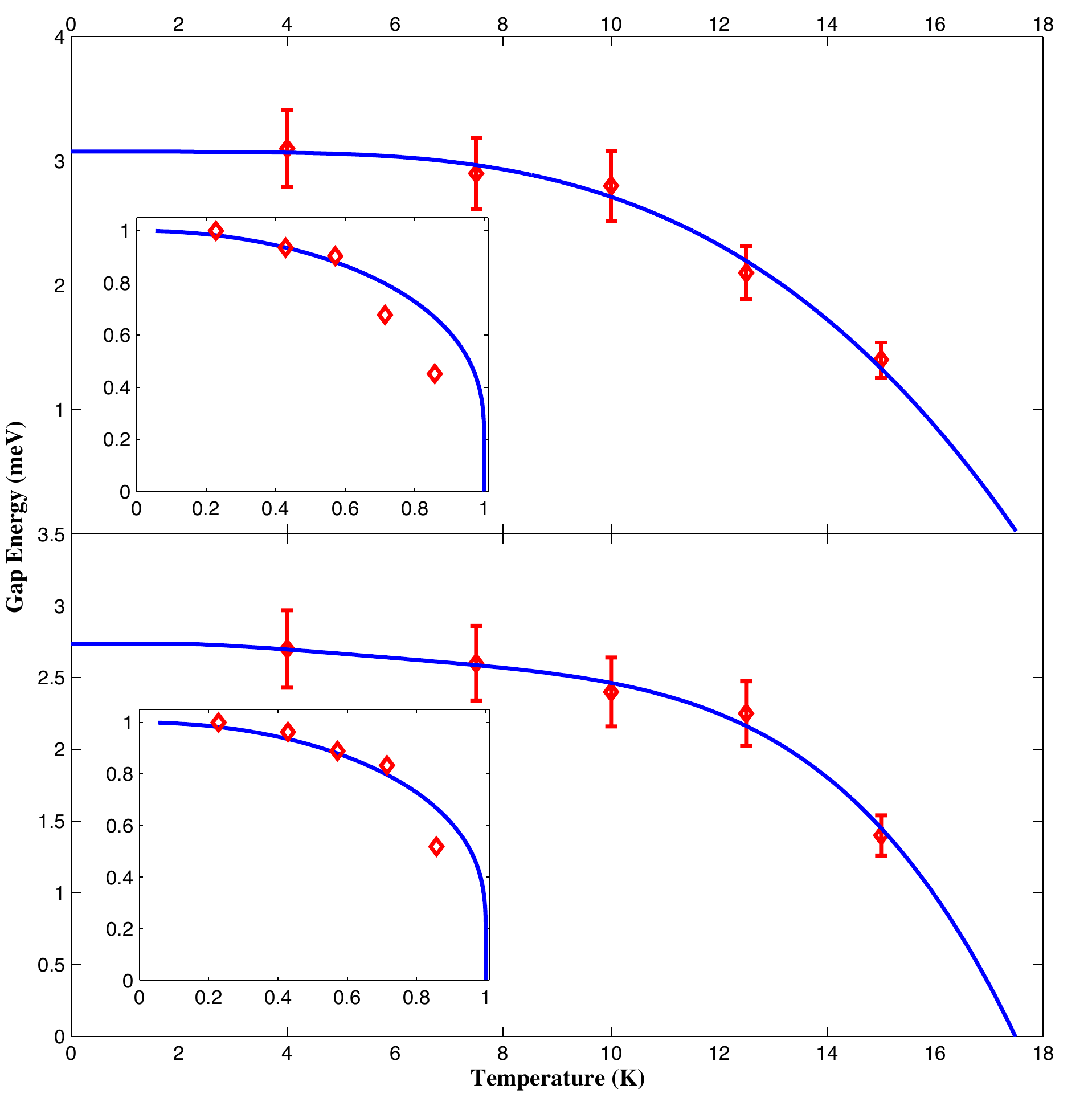}%
\caption{\label{mftgap}The gap value as a function of temperature for the a-axis (top panel) and the larger c-axis gap (bottom panel). Solid lines are guides to the eye to show the trend of the gap values. Insets show the behaviour expected from a mean-field BCS model, from which the data deviates at temperatures close to the transition; at lower temperatures the behaviour is closer to mean-field theory. The smaller c-axis gap is nearly constant to within our sensitivity and is not shown. Error bars are determined from the largest change of gap value that can be made to agree with the data if the other parameters in the model are varied.}
\end{figure}

Figure \ref{mftgap} shows the evolution of the three gaps with temperature, as determined by fitting our model. Neither of the observed gaps whose onset we can detect follow the behaviour expected from a mean field BCS model in the region near the transition, but the fit becomes closer at lower temperatures. It can be noted here that good agreement with BCS theory can be achieved if the transition is assumed to happen at 16 K rather than at 17.5 K, a feature also observed in Aynajian \etal \cite{aynajian}, however transport measurements rule this out and the bulk transition certainly happens at 17.5 K. 

Neutron scattering work \cite{wiebe} shows that a series of incommensurate spin excitations corresponding to the wavevectors (1$\pm$4,0,0) become gapped at the transition, and this must account for a significant amount of the entropy lost. The neutron scattering data reveals that charge and spin degrees of freedom are very strongly coupled, and the Fermi surface reconstruction at $T_{HO}$ has a corresponding effect on the spin excitation spectrum. The energy of the a-axis gap seen in optics (3.2 meV at 4 K) is in good agreement with the gap seen in neutron scattering \cite{wiebe} ($\sim$3.5 meV at 1.5 K). This close correspondence suggests that the spin- and charge-degrees of freedom are strongly correlated, and that the same gap exists in the charge excitation spectrum as in the spin excitation spectrum. 

The neutron scattering results also show commensurate excitations corresponding to the antiferromagnetic zone centre that become gapped at the HO transition. This gap has a value of 2 meV at 1.5 K, very close to the observed value of the smaller gap in the c-axis optical data (1.8 meV at 4 K). The oscillator strength in our model associated with the opening of this gap is much smaller than for the a-axis gap; this is also in agreement with INS which sees the commensurate mode as considerably weaker than the incommensurate modes. This is further good evidence that the gap seen in the optical data is related to the gap in the spin excitation spectrum seen in the neutron scattering.

Recent ARPES \cite{santander} measurements and STM \cite{schmidt, aynajian} studies show the behaviour of the band structure at the transition near the Fermi level. ARPES shows a heavy quasiparticle band that crosses the Fermi level from above at the hidden order transition and hybridizes with a light hole band. Measurements of the differential conductance using STM see a similar effect; a light band breaks into two heavy bands at the transition, forming a gap at the Fermi level. The STM data suggests an effective mass of 5 $m_e$ in the coherence region (in agreement with optical results \cite{nagel}) increasing to $\sim 25$ $m_e$ in the HO state, while ARPES sees an effective mass of 22 $m_e$ in the HO state.

The STM \cite{schmidt, aynajian} show a heavy band splitting at the transition into two heavy bands with a gap of $\sim$4 meV at the Fermi level. However, when the band splits there are additional empty states that appear above the Fermi energy with a gap of approximately 2 meV (see Schmidt \etal \cite{schmidt} Figure 5c, and Aynajian \etal Figure 4c). This band of empty states is available for transitions from the filled band at the Fermi level, and the gap between these two bands shows close agreement with the smaller gap seen in the optical conductivity measured along the c-axis. We therefore suggest that this band accounts for not only the corresponding optically-observed gap, but the commensurate spin excitation gap as well.

The origin of the larger gap seen in the c-axis is less clear: its value at 4 K (2.7 meV) does not correspond well with any gap seen in neutron scattering. APRES measurements \cite{santander} see a heavy quasiparticle band dispersing ~4 meV below $E_F$. STM measurements, likewise, do not see any gaps corresponding to 3 meV between filled states and empty states (though arguably there is a gap between two empty bands of about this magnitude). Both ARPES and STM, however, are limited to measuring a cleaved surface, while we observe this effect only in the c-direction. We suggest that this gap has not been previously observed using the other available probes. The fact that it does not correspond to a gap in the spin excitation spectrum suggests that for the charge carriers involved, spin and charge degrees of freedom are decoupled.

\section{Conclusions} 

We have measured the optical conductivity of the heavy fermion compound URu$_2$Si$_2$ through the hidden order transition at 17.5 K down to 4 K. We observe several s-wave gaps opening in the Fermi surface; one can be seen in the a-axis conductivity and two others in the c-axis. We associate the gap in the a-axis with a value of 3.2 meV at 4 K with the gaps seen by ARPES, STM, and neutron scattering with values of $\sim$ 4 meV, and the smaller gap in the c-axis with a value of 1.8 meV at 4 K with the gapped commensurate spin excitations seen in neutron scattering and the band splitting seen in STM. The larger c-axis gap, with a value of 2.7 meV at 4 K, cannot be associated with any previously reported measured gaps in the magnetic excitation spectrum, and appears to be a purely charge gap. The combination of isotropic s-wave gaps with a gradually narrowing Drude peak provides a complete and consistent explanation for the low frequency optical conductivity in the hidden order state.

We thank  K. Behnia, J.C. Carbotte, A.V.  Chubukov, P Coleman, J.C. Davis, B. Gaulin,  B. Maple, D.L.  Maslov,  A.J. Millis and D.B. Tanner for helpful discussions. In particular we thank K. Behnia for supplying us with unpublished data. This work has been supported by the Natural Science and Engineering Research Council of Canada and the Canadian Institute for Advanced Research. Work in Tallinn was supported by the Estonian Ministry of Education and Research under Grant SF0690029s09, and Estonian Science Foundation under Grants  ETF8170 and ETF8703. The support of Estonian Ministry of Education and Research Grant SF0690029s09, Estonian Science Foundation Grants ETF8170 and ETF8703 is acknowledged.


\begin{thebibliography}{99}

\bibitem{schoenes} J Schoenes, C. Schoenenberger, J. J. M. Franse, and A. A. Menovsky, Phys. Rev. B {\bf 35}, 5375 (1987).

\bibitem{maple} M. B. Maple, J. W. Chen, Y. Dalichaouch, T. Kohara, C. Rossel, M. S. Torikachvili, M. W. McElfresh, and J. D. Thompson, Phys. Rev. Lett. {\bf 56}, 185 (1986).

\bibitem{palstra2} T. T. M. Palstra, A. A. Menovsky, and J. A. Mydosh, Phys. Rev. B {\bf 33}, 6527 (1986).

\bibitem{mcelfresh} M. W. McElfresh, J. D. Thompson, J. O. Willis, M. B. Maple, T. Kohara, and M. S. Torikachvili, Phys. Rev. B {\bf 35}, 43 (1987).

\bibitem{zhu} Z. Zhu, E. Hassinger, Z. Xu, D. Aoki, J. Flouquet, and K. Behnia, Phys. Rev. B {\bf 80} 172501 (2009).

\bibitem{schmidt} A. R. Schmidt, M. H. Hamidian, P. Wahl, F. Meier, A. V. Balatsky, J. D. Garrett, T. J. Williams, G. M. Luke, and J. C. Davis, Nature {\bf 465}, 570 (2010).

\bibitem{nagel} U. Nagel, T. Uleksin, Toomas R{\~o\~o}m, R. P. S. M. Lobo, P. Lejay, C. C. Homes, J. Hall, A. W. Kinross, S. Purdy, T. J. Williams, G. M. Luke, and T. Timusk", arXiv:1107.5574 (2011).

\bibitem{palstra} T. T. M. Palstra, A. A. Menovsky, J. van den Berg, A. J. Dirkmaat, P. H. Kes, G. J. Nieuwenhuys, and J. A. Mydosh, Phys. Rev. Lett. {\bf 55}, 2727 (1985).

\bibitem{schlabitz} W Schlabitz, J. Baumann, R. Pollit, U. Rauchschwalbe, H.M. Mayer, U. Ahlheim, and C. D. Bredl, Z. Phys. B {\bf 55}, 171 (1986).

\bibitem{broholm} C. Broholm, J. K. Kjems, W. J. L. Buyers, P. Matthews, T. T. M. Palstra, A. A. Menovsky, and J. A. Mydosh, Phys. Rev. Lett. {\bf 58}, 1467 (1987).

\bibitem{kiss} Annamária Kiss and Patrik Fazekas, Phys. Rev. B {\bf 71}, 054415 (2005).

\bibitem{chandra} P. Chandra, P. Coleman, J. A. Mydosh, and V. Tripathi, Nature {\bf 417}, 6891 (2002).

\bibitem{haule} Kristjan Haule and Gabriel Kotliar, Nature Phys. {\bf 5}, 796 (2009).

\bibitem{cricchio} Francesco Cricchio, Fredrik Bultmark, Oscar Granas, and Lars Nordstrom, Phys. Rev. Lett. {\bf 103}, 107202 (2009).

\bibitem{harima} Hisatomo Harima, Kazumasa Miyake, and Jacques Flouquet, Journal of the Physical Society of Japan {\bf 79}, 033705 (2010).

\bibitem{bonn}  D. A. Bonn, J. D. Garrett, and T. Timusk, Phys. Rev. Lett. {\bf 61}, 1305 (1988).

\bibitem{degiorgi}  L. Degiorgi, St. Thieme, H. R. Ott, M. Dressel, G. Gruner, Y. Dalichaouch, M. B. Maple, Z. Fisk, C. Geibel, and F. Steglich, Z Phys.B {\bf 102}, 367 (1996).

\bibitem{crazy} J. Levallois, F. Levy-Bertrand, M. K. Tran, D. Stricker, J. A. Mydosh, Y.-K. Huang, and D. van der Marel, Phys. Rev.B {\bf 84}, 4420 (2011).

\bibitem{homes} C.C. Homes, M.A. Reedyk, D.A. Crandles, and T. Timusk, Applied Optics {\bf 32}, 2976 (1993).

\bibitem{dynes}  R. C. Dynes, V. Narayanamurti, and J. P. Garno, Phys. Rev. Lett. {\bf 41}, 1509 (1978).

\bibitem{mydoshrev} J. A. Mydosh and P. M. Oppeneer, arXiv:1107.0258 (2011).


\bibitem{santander} A. F. Santander-Syro, M. Klein, F. L. Boariu, A. Nuber, P. Lejay, and F. Reinert", Nature Phys. {\bf 5}, 637 (2009).

\bibitem{aynajian} P. Aynajian, E. H. da Silva Neto, C. V. Parker, Y.-K. Huang, A. Pasupathy, J. A. Mydosh, and A. Yazdani, Proc. Natl. Acad. USA, {\bf 107}, 10383 (2010).

\bibitem{wiebe} C. R. Wiebe, J. A. Janik, G. J. MacDougall, G. M. Luke, J. D. Garrett, H. D. Zhou, Y.-J. Jo, L. Balicas, Y. Qiu, J. R. D. Copley, Z. Yamani, and W. J. L. Buyers, Nature Phys. {\bf 3}, 96 (2007).

\end{thebibliography}
\end{document}